\documentstyle[12pt]{article}

\begin{document}

\title{An effective formulation on quantum hadrodynamics \\
at finite temperatures and densities}

\author{{B. X. Sun$^{1}$\thanks{Corresponding author.
E-mail address: sunbx@bjut.edu.cn. The project supported by the
Foundations of Beijing University of Technology and
Ren-Cai-Qiang-Jiao Foundation of Beijing Municipal Education
Commission}, X. F. Lu$^{2}$, L. Li$^{3}$, P. Z. Ning$^{3}$,
P. N. Shen$^{4}$, E. G. Zhao$^{5}$}\\
\normalsize{${}^{1}$College of Applied Sciences, Beijing University
of Technology, Beijing 100022, China} \\
\normalsize{${}^{2}$Department of Physics, Sichuan University,
Chengdu  610064, China} \\
\normalsize{${}^{3}$Department of Physics, Nankai University,
Tianjin  300071, China} \\
\normalsize{${}^{4}$Institute of High Energy Physics, The Chinese
Academy of Sciences,} \\
\normalsize{P.O.Box 918(4), Beijing  100039, China} \\
\normalsize{${}^{5}$Institute of Theoretical Physics, The Chinese
Academy of Sciences, Beijing  100080, China} \\
}

\maketitle

\begin{abstract}
According to Wick's theorem, the second order self-energy
corrections of hadrons in the hot and dense nuclear matter are
calculated. Furthermore, the Feynman rules are summarized, and an
effective formulation on quantum hadrodynamics at finite
temperatures and densities is evaluated. As the strong couplings
between nucleons are considered, the self-consistency of this method
is discussed in the framework of relativistic mean-field
approximation. Debye screening masses of the scalar and vector
mesons in the hot and dense nuclear matter are calculated with this
method in the relativistic mean-field approximation. The results are
different from those of thermofield dynamics and Brown-Rho
conjecture. Moreover, the effective masses of the photon and the
nucleon in the hot and dense nuclear matter are discussed.
\end{abstract}



\newpage

\section{Introduction}
The quantum field theory, introduced to solve the many-body problems
since the late fifties, has proved to be highly successful in
studying the ground state, the equilibrium and non-equilibrium
properties at finite
temperatures\cite{Wa.74,Chin.77,WS.86,Chou.85,Ume.82,Kapu.89,Furn.91,Bell.96,Das.97}.
These theories to solve many-body problems are based on the
conjecture that all of the effects of the medium at finite
temperatures and densities would change the propagators of
particles. When the propagators in the mediums are obtained, the
properties of particles and the equation of state of the medium can
be calculated easily. The propagators of particles in the medium are
different from those in vacuum correspondingly, so this conjecture
means the redefinition of
 the {\sl vacuum}, and the calculation procedures are all
carried out in this new {\sl vacuum}.

In a series of our previous papers\cite{Sun.02,Sun.0206,Sun.0207},
according to Wick's theorem, we have evaluated an effective
formulation on quantum hadrodynamics to solve the nuclear many-body
problems. In this formulation, the Feynman propagators in vacuum are
adopted, and the second order self-energies of particles in the
nuclear matter are calculated, while the effects of the nuclear
matter are treated as the condensations of nucleons and mesons. It
shows the same results as the method of quantum
hadrodynamics\cite{Wa.74,Chin.77,WS.86}. With this new method, we
have studied the effective masses of the photon and
mesons\cite{Sun.02,Sun.0206}, and constructed the density-dependent
relativistic mean-field model\cite{Sun.0206}. In this paper we will
generalize this new method to the situation at finite temperatures
and densities.

The organization of this paper is as follows. The second order
self-energy corrections of the nucleon and the meson in the nuclear
matter are calculated from Wick expansion in the framework of
quantum hadrodynamics 1 in Sec.~\ref{sect:self}, then the Feynman
rules and self-consistency for this method at finite temperatures
and densities are summarized in Sec.~\ref{Feynman rules}. The
results of screening masses of mesons are presented in
Sec.~\ref{Debye}. In Sec.~\ref{Photon}, the effective mass of the
photon in the hot nuclear matter is discussed. The summary is given
in Sec.~\ref{Summary}.

\section{The self-energies of hadrons in the hot and dense nuclear matter}
\label{sect:self}

 According to Walecka-1 model, the Lagrangian density in
 nuclear matter can be written as
\begin{eqnarray}
\label{eq:Lar}
{\cal L}~&=&~\bar\psi \left(i\gamma_{\mu}\partial^{\mu}  -
M_N\right)\psi~+~\frac{1}{2}\partial_\mu\sigma\partial^\mu\sigma-\frac{1}{2}
m^2_\sigma \sigma^2_{}
-\frac{1}{4}\omega_{\mu\nu}\omega^{\mu\nu}+ \frac{1}{2}
m^2_\omega\omega_\mu\omega^\mu \nonumber \\
&&-g_\sigma\bar\psi\sigma\psi-g_\omega\bar\psi \gamma_\mu
\omega^\mu \psi,
\end{eqnarray}
where $\psi$ is the field of the nucleon, $M_N$ is the
nucleon mass, and
\begin{equation}
\omega_{\mu\nu}~=~\partial_\mu\omega_\nu-\partial_\nu\omega_\mu,
\end{equation}
is the field tensor of the vector meson.

In the hot nuclear matter, the expectations of
normal products of the creation
 operator and corresponding annihilation operator are relevant to the
 distribution functions of particles.
\begin{eqnarray}
\langle~N,\beta~|A_{p^{\prime}\lambda^{\prime}}^{\dagger}A_{p\lambda}|~N,\beta~\rangle&=&n_F
\delta^3(\vec{p^\prime}-\vec{p})\delta_{\lambda^\prime \lambda},
\\
\langle~N,\beta~|B_{p^{\prime}\lambda^{\prime}}^{\dagger}B_{p\lambda}|~N,\beta~\rangle
&=&\bar n_F\delta^3(\vec{p^\prime}-\vec{p})\delta_{\lambda^\prime
\lambda},
\end{eqnarray}
where $\lambda^\prime$ and $\lambda$ denote spins of fermions,
$n_F$ and $\bar n_F$ are the distribution functions of the
nucleon and antinucleon, respectively.
\begin{eqnarray}
n_F&=&\frac{1}{\exp{[(E(p)-\mu)/T]}+1}, \\
\bar n_F&=&\frac{1}{\exp{[(E(p)+\mu)/T]}+1},
\end{eqnarray}
where $E(p)~=~\sqrt{\vec{p}^2~+~M_N^2}$, and $\mu$ is the chemical
potential of the nucleon. The relation of chemical
potential and the number density of nucleons $\rho_B$ is
\begin{equation}
\label{eq:vmidu} \rho_B~=~\frac{2}{(2\pi)^3}\int d^3p\left[n_F-\bar
n_F \right],
\end{equation}

\begin{eqnarray}
\langle~N,\beta~|a_{k^{\prime}}^{\dagger}a_{k}|~N,\beta~\rangle&=&
n_\sigma \delta^3(\vec{k^\prime}-\vec{k}),\\
\langle~N,\beta~|b_{k^{\prime}\delta^{\prime}}^{\dagger}b_{k\delta}|~N,\beta~\rangle&=&
n_\omega \delta^3(\vec{k^\prime}-\vec{k})\delta_{\delta^\prime
\delta},
\end{eqnarray}
where $\delta^\prime$ and $\delta$ denote the spins of the vector meson,
$n_\sigma$ and $n_\omega$ are the boson distribution
functions of the scalar and vector mesons, respectively.
\begin{equation}
\label{eq:boson}
n_\alpha=\frac{1}{\exp{[|\Omega_\alpha|/T}-1]},~~~~\alpha~=~\sigma,~~\omega,
\end{equation}
where $\Omega_\alpha=\sqrt{\vec{k}^2+m^2_\alpha}$.
Since the meson number is not conserved in nuclear matter, there
is not the contribution of the meson chemical potential in the
boson distribution function of Eq.~(\ref{eq:boson}).

The finite-temperature state of nuclear matter could be understood
as the state that there are a great number of coupling nucleons, antinucleons
and mesons in the perturbation vacuum, so the noninteracting
propagators in vacuum are used in the calculation of the
self-energies of particles.

The momentum-space noninteracting propagators of the scalar meson,
vector meson and nucleon in perturbation vacuum $|~0~\rangle$
follow as:
\begin{equation}
\label{eq:scalar} i\Delta_0(p) =
\frac{-1}{p^2-m^2_\sigma+i\varepsilon},
\end{equation}
\begin{equation}
 \label{eq:vector}
iD_0^{\mu\nu}(p) =
\frac{g^{\mu\nu}}{p^2-m^2_\omega+i\varepsilon},
\end{equation}
\begin{equation}
\label{eq:baryon} iG_0^{\alpha\beta}(p) = \frac{-1}{\gamma_\mu
p^\mu-M_N+i\varepsilon}.
\end{equation}
Since the vector meson couples to the conserved baryon current, the
longitudinal part in the propagator of the vector meson will not contribute
to physical quantities\cite{WS.86,Bj.65}. Therefore, only the transverse
part in the propagator of the vector meson is written
in Eq.~(\ref{eq:vector}).

According to Eq.~(\ref{eq:Lar}), the interaction Hamiltonian can be
expressed as
\begin{equation}
{\cal H}_I~=~ g_\sigma\bar\psi\sigma\psi+g_\omega\bar\psi
\gamma_\mu \omega^\mu \psi.
\end{equation}
In the second order approximation, only
\begin{equation}
\hat{S}_2 ~=~\frac{(-i)^2}{2 !}\int d^{4}x_1 \int d^{4}x_2
  T \left[{\cal H}_I(x_1){\cal H}_I(x_2) \right]
\end{equation}
in the S-matrix
should be calculated in order to obtain the self-energy corrections of the
nucleon and mesons.

The coupling constants $g_\sigma$, $g_\omega$, the masses of
mesons $m_\sigma$, $m_\omega$ and the nucleon mass $M_N$ are
supposed to have been renormalized, then the contributions of the
nucleon-loop diagrams are not needed to be
considered\cite{Bj.65,Bj.64}. Only the terms with one contraction of two
fields in the Wick expansions of $\hat{S}_2$ should be calculated.
Thus with the similar procedure of
Ref.\cite{Sun.02,Sun.0206,Sun.0207}, the scalar
meson self-energy $\Sigma_{\sigma}$, the vector meson
self-energy $\Sigma_{\omega}$, the nucleon self-energy
$\Sigma^{+}$ and the antinucleon self-energy $\Sigma^{-}$ in the hot and
dense nuclear matter
 are obtained.

The second order self-energy of the scalar meson is
\begin{eqnarray}
\label{eq:scalar-self} \Sigma_\sigma&=&
(-ig_\sigma)^2~\sum_{\lambda=1,2} \int\frac{d^3p}{(2\pi)^3}
\frac{M_N}{E(p)} \nonumber \\
&&\left[n_F \bar{U}(p,\lambda) \left(iG_0(p-k)+iG_0(p+k)\right)
 U(p,\lambda)\right. \nonumber \\
&&-\left.\bar n_F \bar{V}(p,\lambda) \left(iG_0(-p-k)+iG_0(-p+k)\right)
V(p,\lambda)\right] \nonumber \\
&=&
g_\sigma^2~\sum_{\lambda=1,2} \int\frac{d^3p}{(2\pi)^3}
\frac{M_N}{E(p)} \nonumber \\
&&\left[n_F \bar{U}(p,\lambda) \left(\frac{1}{\rlap{/} p -\rlap{/}
k - M_N}~+~ \frac{1}{\rlap{/} p +\rlap{/} k - M_N}\right)
 U(p,\lambda)\right. \nonumber \\
&&-\left.\bar n_F \bar{V}(p,\lambda) \left(\frac{1}{-\rlap{/} p
-\rlap{/} k - M_N}~+~ \frac{1}{-\rlap{/} p +\rlap{/} k -
M_N}\right) V(p,\lambda)\right],
\end{eqnarray}
where $U(p,\lambda)$ and $V(p,\lambda)$ are the Dirac spinors of the
nucleon and the antinucleon, respectively, and $\bar{U}(p,\lambda)$
and $\bar{V}(p,\lambda)$ are their conjugate spinors, respectively.
\begin{equation}
\label{eq:nor1}
\sum_{\lambda=1,2}U(p,\lambda)\bar U(p,\lambda)
=\frac{\rlap {/} p~+~M_N}{2M_N},
\end{equation}
\begin{equation}
\label{eq:nor2}
\sum_{\lambda=1,2}-V(p,\lambda)\bar V(p,\lambda)
=\frac{-\rlap {/} p~+~M_N}{2M_N}.
\end{equation}

The second order self-energy of the vector meson is
\begin{eqnarray}
\label{eq:vector-self} -g_{\mu\nu}~\Sigma_\omega&=&
(-ig_\omega)^2~\sum_{\lambda=1,2}\int\frac{d^3p}{(2\pi)^3}
\frac{M_N}{E(p)} \nonumber \\
&&\left[n_F \bar{U}(p,\lambda) \left(\gamma_{\nu}iG_0(p-k)\gamma_{\mu}~+~ \gamma_{\mu}iG_0(p+k)
\gamma_{\nu}\right) U(p,\lambda) \right. \nonumber \\
&&- \left.\bar n_F \bar{V}(p,\lambda)
\left(\gamma_{\nu}iG_0(-p-k)\gamma_{\mu}~+~ \gamma_{\mu}iG_0(-p+k)\gamma_{\nu}\right)
 V(p,\lambda) \right] \nonumber \\
&=&
{g_\omega}^2~\sum_{\lambda=1,2}\int\frac{d^3p}{(2\pi)^3}
\frac{M_N}{E(p)} \nonumber \\
&&\left[n_F \bar{U}(p,\lambda) \left(\gamma_{\nu}\frac{1}{\rlap{/}
p -\rlap{/} k - M_N}\gamma_{\mu}~+~ \gamma_{\mu}\frac{1}{\rlap{/}
p +\rlap{/} k - M_N}\gamma_{\nu}\right)
 U(p,\lambda) \right. \nonumber \\
&&- \left.\bar n_F \bar{V}(p,\lambda)
\left(\gamma_{\nu}\frac{1}{-\rlap{/} p -\rlap{/} k -
M_N}\gamma_{\mu}~+~ \gamma_{\mu}\frac{1}{-\rlap{/} p +\rlap{/} k -
M_N}\gamma_{\nu}\right)
 V(p,\lambda) \right].
\end{eqnarray}

The second order self-energy of the nucleon is
\begin{equation}
\label{eq:self}
\Sigma^{+}~=~\sum_{s=\sigma,\omega}\left(\Sigma_{s,1}^{+}~+~\Sigma_{s,2}^{+}~+~\Sigma_{s,3}^{+}
\right),
\end{equation}
in which
\begin{eqnarray}
\Sigma_{\sigma,1}^{+}&=&(-ig_\sigma)^2 \sum_{\lambda=1,2}\int
\frac{d^3p}{(2\pi)^3} \frac{M_N}{E(p)}~i\Delta_0(0) \nonumber \\
&&\left[n_F \bar{U}(p,\lambda)U(p,\lambda)
-\bar{n}_F \bar{V}(p,\lambda)V(p,\lambda)\right] \nonumber \\
&=&- \frac{g^2_\sigma}{m^2_\sigma} \rho_S,
\end{eqnarray}
where $\rho_S$ is the scalar density of protons or neutrons,
\begin{equation}
\label{eq:biaol}
\rho_S~=~\frac{2}{(2\pi)^3} \int d^3p
\frac{M_N}{\sqrt{\vec{p}^2+{M_N}^2}}\left(n_F+\bar n_F \right),
\end{equation}
\begin{eqnarray}
\Sigma_{\sigma,2}^{+}&=&g^2_\sigma \sum_{\lambda=1,2}\int \frac{d^3p}{(2\pi)^3}
\frac{M_N}{E(p)} \nonumber \\
&&\left[n_F U(p,\lambda)~i\Delta_0(k-p)~\bar{U}(p,\lambda)
- \bar n_F V(p,\lambda)~i\Delta_0(k+p)~\bar{V}(p,\lambda)
\right]  \\
&=&-g^2_\sigma \int \frac{d^3p}{(2\pi)^3}
\frac{M_N}{E(p)}
\left[n_F \frac{\rlap{/} p+M_N}{2M_N}
\frac{1}{(k-p)^2-m^2_\sigma} - \bar n_F \frac{\rlap{/}
p-M_N}{2M_N} \frac{1}{(k+p)^2-m^2_\sigma}\right], \nonumber
\end{eqnarray}
\begin{eqnarray}
\Sigma_{\sigma,3}^{+}&=&(-ig_\sigma)^2 \int
\frac{d^3k}{2\Omega_\sigma (2\pi)^3}
~n_\sigma \left(iG_0(p-k)+iG_0(p+k)\right) \nonumber \\
&=&g^2_\sigma \int
\frac{d^3k}{2\Omega_\sigma (2\pi)^3}
~n_\sigma \left(\frac{1}{\rlap{/} p-\rlap{/}
k-M_N}+\frac{1}{\rlap{/} p+\rlap{/} k-M_N} \right),
\end{eqnarray}
\begin{eqnarray}
\Sigma_{\omega,1}^{+}&=&(-ig_\omega)^2 \sum_{\lambda=1,2}\int
\frac{d^3p}{(2\pi)^3} \frac{M_N}{E(p)}\gamma_\mu~iD_0^{\mu\nu}(0) \nonumber \\
&&\left[n_F \bar{U}(p,\lambda)\gamma_\nu U(p,\lambda)
-\bar{n}_F \bar{V}(p,\lambda)\gamma_\nu V(p,\lambda)\right] \nonumber \\
&=&\gamma_0 \frac{g^2_\omega}{m^2_\omega}
\rho_B,
\end{eqnarray}
where $\rho_B$ defined in Eq.~(\ref{eq:vmidu}) is the number density
of protons or neutrons, and
\begin{eqnarray}
\Sigma_{\omega,2}^{+}&=&g^2_\omega \sum_{\lambda=1,2}\int \frac{d^3p}{(2\pi)^3}
\frac{M_N}{E(p)} \nonumber \\
&&\left[n_F \gamma_\mu U(p,\lambda)~iD_0^{\mu\nu}(k-p)~\bar{U}(p,\lambda) \gamma_\nu
- \bar n_F \gamma_\mu V(p,\lambda)~iD_0^{\mu\nu}(k+p)~\bar{V}(p,\lambda) \gamma_\nu
\right] \nonumber \\
&=&g^2_\omega \int \frac{d^3p}{(2\pi)^3}
\frac{M_N}{E(p)} \nonumber \\
&&\left(n_F \gamma_\mu \frac{\rlap{/} p+M_N}{2M_N} \gamma_\nu
\frac{g^{\mu\nu}}{(k-p)^2-m^2_\omega} - \bar n_F \gamma_\mu
\frac{\rlap{/} p-M_N}{2M_N} \gamma_\nu
\frac{g^{\mu\nu}}{(k+p)^2-m^2_\omega}\right),
\end{eqnarray}
\begin{eqnarray}
\Sigma_{\omega,3}^{+}&=&(-ig_\omega)^2 \sum_{\delta=1,2,3} \int \frac{d^3k}{2\Omega_\omega(2\pi)^3}
n_\omega \gamma^\mu \varepsilon_\mu(k,\delta)\left[iG_0(p-k)+iG_0(p+k)\right]
\gamma^\nu \varepsilon_\nu(k,\delta) \nonumber \\
&=&g_\omega^2 \sum_{\delta=1,2,3} \int \frac{d^3k}{2\Omega_\omega(2\pi)^3}
n_\omega \gamma^\mu \varepsilon_\mu(k,\delta)\left[\frac{1}{\rlap{/} p-\rlap{/}
k-M_N}+\frac{1}{\rlap{/} p+\rlap{/} k-M_N} \right]
\gamma^\nu \varepsilon_\nu(k,\delta), \nonumber \\
\end{eqnarray}
where $\varepsilon_\mu(k,\delta)$ and $\varepsilon_\nu(k,\delta)$
are the space-like orthonormalized vectors of the vector meson.

The second order self-energy of the antinucleon in the nuclear matter is
\begin{equation}
\Sigma^{-}~=~\sum_{s=\sigma,\omega}\left(\Sigma_{s,1}^{-}~+~\Sigma_{s,2}^{-}~+~\Sigma_{s,3}^{-}
\right),
\end{equation}
in which
\begin{equation}
\Sigma_{\sigma,1}^{-}~=~-\Sigma_{\sigma,1}^{+}~=~\frac{g^2_\sigma}{m^2_\sigma} \rho_S,
\end{equation}
\begin{eqnarray}
\Sigma_{\sigma,2}^{-}&=&
-g^2_\sigma \sum_{\lambda=1,2}\int \frac{d^3p}{(2\pi)^3}
\frac{M_N}{E(p)} \nonumber \\
&&\left[n_F U(p,\lambda)~i\Delta_0(-k-p)~\bar{U}(p,\lambda)
- \bar n_F V(p,\lambda)~i\Delta_0(-k+p)~\bar{V}(p,\lambda)
\right] \\
&=&g^2_\sigma \int \frac{d^3p}{(2\pi)^3}
\frac{M_N}{E(p)}
 \left(n_F \frac{\rlap{/} p+M_N}{2M_N}
\frac{1}{(-k-p)^2-m^2_\sigma} - \bar n_F \frac{\rlap{/}
p-M_N}{2M_N} \frac{1}{(-k+p)^2-m^2_\sigma}\right), \nonumber
\end{eqnarray}
\begin{eqnarray}
\Sigma_{\sigma,3}^{-}&=&-(-ig_\sigma)^2 \int
\frac{d^3k}{2\Omega_\sigma (2\pi)^3}
~n_\sigma \left(iG_0(-p-k)+iG_0(-p+k)\right) \nonumber \\
&=&-g^2_\sigma \int
\frac{d^3k}{2\Omega_\sigma (2\pi)^3}
~n_\sigma \left(\frac{1}{-\rlap{/} p-\rlap{/}
k-M_N}+\frac{1}{-\rlap{/} p+\rlap{/} k-M_N} \right),
\end{eqnarray}
\begin{equation}
\Sigma_{\omega,1}^{-}~=~-\Sigma_{\omega,1}^{+}~=~-\gamma_0 \frac{g^2_\omega}{m^2_\omega}
\rho_B,
\end{equation}
\begin{eqnarray}
\Sigma_{\omega,2}^{-}&=&-g^2_\omega \sum_{\lambda=1,2}\int \frac{d^3p}{(2\pi)^3}
\frac{M_N}{E(p)} \nonumber \\
&&\left[n_F \gamma_\mu U(p,\lambda)~iD_0^{\mu\nu}(-k-p)~\bar{U}(p,\lambda) \gamma_\nu
- \bar n_F \gamma_\mu V(p,\lambda)~iD_0^{\mu\nu}(-k+p)~\bar{V}(p,\lambda) \gamma_\nu
\right] \nonumber \\
&=&-g^2_\omega \int \frac{d^3p}{(2\pi)^3}
\frac{M_N}{E(p)} \nonumber \\
&&\left(n_F \gamma_\mu \frac{\rlap{/} p+M_N}{2M_N} \gamma_\nu
\frac{g^{\mu\nu}}{(-k-p)^2-m^2_\omega} - \bar n_F \gamma_\mu
\frac{\rlap{/} p-M_N}{2M_N} \gamma_\nu
\frac{g^{\mu\nu}}{(-k+p)^2-m^2_\omega}\right),
\end{eqnarray}
\begin{eqnarray}
\Sigma_{\omega,3}^{-}&=&-(-ig_\omega)^2 \sum_{\delta=1,2,3} \int \frac{d^3k}{2\Omega_\omega(2\pi)^3}
n_\omega \gamma^\mu \varepsilon_\mu(k,\delta)\left[iG_0(-p-k)+iG_0(-p+k)\right]
\gamma^\nu \varepsilon_\nu(k,\delta) \nonumber \\
&=&-g_\omega^2 \sum_{\delta=1,2,3} \int \frac{d^3k}{2\Omega_\omega(2\pi)^3}
n_\omega \gamma^\mu \varepsilon_\mu(k,\delta)\left[\frac{1}{-\rlap{/} p-\rlap{/}
k-M_N}+\frac{1}{-\rlap{/} p+\rlap{/} k-M_N} \right]
\gamma^\nu \varepsilon_\nu(k,\delta). \nonumber \\
\end{eqnarray}

\section{Feynman rules}
\label{Feynman rules}

With the propagators in vacuum as Eqs.~(\ref{eq:scalar}), ~(\ref{eq:vector}) and
~(\ref{eq:baryon}), the second order self-energies of particles in
the hot nuclear matter are obtained. Therefore,
the expectation values of observables built out of products of
field operators are allowed to be calculated. For the present
method, the Feynman rules are as follows:

1. Draw all topologically distinct, connected diagrams.

2. Include the following factors for the scalar and vector
vertices respectively,
\begin{equation}
-ig_\sigma;~~~~~scalar~~~~~~~~~~~~~-ig_\omega
\gamma_\mu;~~~~~vector.
\end{equation}

3. Include the factors for the scalar, vector, and baryon
propagators as Eqs.~(\ref{eq:scalar}), ~(\ref{eq:vector}) and
~(\ref{eq:baryon}), respectively.

4. Conserve four-momentum at each vertex, and include a factor of
$(2\pi)^4 \delta(\sum p)$ at each vertex.

5. Include a factor of
$$\sum_{\lambda=1,2}\int\frac{d^3p}{(2\pi)^3}
\frac{M_N}{E(p)}n_F$$for each pair of crosses for external lines
of the nucleon.

6. Include a factor of $$\sum_{\lambda=1,2}\int\frac{d^3p}{(2\pi)^3}
\frac{M_N}{E(p)}(-1) \bar n_F$$for each pair of crosses for
external lines of the antinucleon.

7. Include a factor of $$ \int d^3 \tilde{k}~ n_\alpha$$
for each pair of crosses for external lines of the scalar or vector meson,
where
$$ \int d^3 \tilde{k}~ n_\alpha ~=~\left\{
\begin{array}{rr}
 \int \frac{d^3k}
{2\Omega_\sigma(2\pi)^3}n_\sigma,&
~~\alpha=\sigma,\\
\sum\limits_{\delta=1,2,3} \int \frac{d^3k}
{2\Omega_\omega(2\pi)^3}n_\omega,&
~~\alpha=\omega,
\end{array}
\right.
$$
and $\Omega_\alpha~
=~\sqrt{\vec{k}^2~+~m^2_\alpha}$.

8. Take the Dirac matrix product along a fermion line.

9. The momentums and spins of external lines with a cross or
without a cross take the same values with each other, respectively.

10. Change the momentum $p$ of the external lines of the nucleon
into $(-p)$, the Dirac spinor $U(p,\lambda)$ and the corresponding
conjugate spinor $\bar U(p,\lambda)$ for the external lines of the
nucleon into the Dirac spinor $V(p,\lambda)$ and the corresponding
conjugate spinor $\bar V(p,\lambda)$ of the antinucleon,
respectively,
\begin{equation}
\label{eq:trans1}
p~\rightarrow~-p,
\end{equation}
\begin{equation}
\label{eq:trans2}
U(p,\lambda)~\rightarrow~V(p,\lambda),
\end{equation}
\begin{equation}
\label{eq:trans2}
\bar U(p,\lambda)~\rightarrow~\bar V(p,\lambda).
\end{equation}
Therefore, the results of corresponding topological diagrams with external
lines of the antinucleon are obtained.

11. Include a factor of $(-1)$ for the second order self-energy of
the antinucleon.

12. Include a factor of $(-1)$ in the calculation of exchange diagrams.

13. Include a factor of $\delta_{ij}$ along a fermion line for
isospin(here $i,j=p,n$).

If the masses of the scalar meson, vector meson, baryon , and all
coupling constants are renormalized, the loop diagrams are not
needed to be considered at all. Therefore, only the diagrams with
crosses should be considered in the calculation of self-energies of
particles.

In Fig.~1, the Feynman diagrams on the second order self-energy of the
meson interaction with nucleons in the hot and dense nuclear matter
are shown. The
Feynman diagrams on the self-energy of the meson interaction with
antinucleons in the nuclear matter can be obtained
by changing the directions of all the fermion lines in each diagram,
respectively.

The Feynman diagrams on the second order self-energy of
the nucleon in the hot and dense nuclear matter
are shown in Fig.~2. According to Eqs.~(\ref{eq:trans1} -
\ref{eq:trans2}), the second order self-energy of the antinucleon in
the hot and dense nuclear matter can be obtained easily.
As the anti-commutation relations are considered, a
factor of $(-1)$ should be added for the second order self-energy
of the antinucleon.

Because of the strong interaction between nucleons,
the coupling constants are very large and the perturbation calculation does
not converge. Although the second-order self-energy of the particle
can be summed to all orders with Dyson equation, This procedure is not
self-consistent, however, since the propagating particle interacts to
all orders, whereas the background particles are noninteracting.
Self-consistency can be achieved by using the interacting propagators
to also determine the self-energy\cite{WS.86}.

In the relativistic mean-field approximation, the meson fields operators can
be replaced by their expectation values in the nuclear matter
\cite{WS.86}:
\begin{eqnarray}
\sigma&\rightarrow&\langle\sigma\rangle~=~\sigma_0, \\
\omega_\mu&\rightarrow&\langle\omega_\mu\rangle~=~\omega_0\delta^{0}_{\mu}.
\end{eqnarray}
Therefore, the off-shell part of the nucleon propagator can be
written as\cite{WS.86,Sun.0206}:
\begin{equation}
\label{eq:barrmf} G^H_{\alpha\beta}(p) = \frac{i}{\gamma_\mu
 p^\mu-M^\ast_N+i\varepsilon},
\end{equation}
where $p~=~(E^\ast(p),\vec{p})$, and
$E^\ast(p)=\sqrt{\vec{p}^2~+~{M^{\ast}_N}^2}$.
$M^{\ast}_N=M_N+g_\sigma\sigma_0$
is the effective mass of the nucleon in the nuclear matter.

With the nucleon propagator in Eq.~(\ref{eq:barrmf}), the self-energies
of particles can be calculated. It corresponds to the transformation of
\begin{equation}
\label{eq:con-rep}
E(p)~\rightarrow~E^{\ast}(p), ~~~~~~
M_N ~\rightarrow~M^{\ast}_N, ~~~~~~
\mu~\rightarrow~\mu~-~
g_\omega \omega_0
\end{equation}
in calculations of Sec.~\ref{sect:self}.

\section{Debye screening effect in the hot and dense nuclear matter}
\label{Debye}

The properties of hadrons in the nuclear matter have caused more attention
of nuclear physicists in the past years\cite{Furn.88,Br.91,Hats.92,Song.93,Jean.94,SRK.95}.
Suppose the momentum of the meson is zero,  the
screening masses of the scalar and vector mesons
in the hot and dense nuclear matter
are obtained from Eqs.~(\ref{eq:scalar-self}) and
~(\ref{eq:vector-self}), respectively.

\begin{equation}
\label{eq:58}
m^\ast_\alpha=\sqrt{m^2_\alpha~+~{\Sigma^\prime_\alpha}},~~~~~
\alpha~=~\sigma, \omega
\end{equation}
in which
\begin{equation}
{\Sigma^\prime_\sigma}
~=~\frac{g_\sigma^2 (\rho_S^p~+~\rho_S^n)}{M^{\ast}_N}, ~~~~~
{\Sigma^\prime_\omega}~=~\frac{g_\omega^2
(\rho_S^p~+~\rho_S^n)}{2M^{\ast}_N},
\end{equation}
where $\rho_S^p$ and $\rho_S^n$ are the scalar densities of protons and
neutrons, respectively.
In the limit of zero momentum of the meson,
 our results on the self-energies of the mesons at zero
temperature are same as those in Ref.\cite{Furn.88}.

With the parameters fixed in Ref.~\cite{HS.81},
and supposing the masses of the scalar and vector meson have been
fixed already,
\begin{equation}
m_\sigma~=~520.0MeV,~~~~~
m_\omega~=~783.0MeV,
\end{equation}
the screening meson masses, the effective mass of the nucleon and
the energy per nucleon in the nuclear matter under different
densities and temperatures are calculated self-consistently in the
framework of the relativistic mean-field approximation.

The screening masses of $\sigma$ and $\omega$ mesons
$m^\ast_\sigma$, $m^\ast_\omega$, the effective mass of the nucleon
$ M^\ast_N$ and the energy per nucleon $E/A$ in the nuclear matter
as functions of the nucleon number density $\rho_N$ with the
temperature $T=10.0MeV$  are displayed in Table 1. We find that the
screening masses of $\sigma$ and $\omega$ mesons both increase with
the nucleon number density, while the effective mass of the nucleon
decreases at the fixed temperature. It implies that the Debye
screening effect enhanced in the denser nuclear matter.

The properties of particles vs temperature at
$\rho_N=\rho_0=0.148fm^{-3}$ are listed in Table 2, where $\rho_0$ is the
saturation density of nuclear matter at zero temperature. The energy per
nucleon increases with the temperature, in other words, the
binding energy decreases when the temperature becomes higher.
Meanwhile, the effective mass of the nucleon increases with
temperature and the screening masses of the scalar and vector
mesons both decrease when the temperature increases at the
fixed density of nuclear matter. Obviously, when the temperature
increases, the Debye screening effect decreases, and the interaction
between nucleons becomes stronger.
Our results are different from those with the thermo-field
dynamics\cite{SRK.95} and the familiar Brown-Rho
conjecture\cite{Br.91}.

Since the screening masses of $\sigma$ meson, $\omega$ meson and the
effective mass of the nucleon are all relevant to the scalar density
of nucleons, where the relativistic effect of the nucleon is
included, it is not difficult to understand the different changes of
them with the temperature and density of the nuclear matter. When
the temperature of the nuclear matter is not too high, i.e.
$M_N>>T$, the distribution function of the nucleon is equal to 1 and
the distribution function of the antinucleon is zero approximately,
\begin{equation}
n_F=1,~~~~~~\bar n_F=0. \nonumber
\end{equation}
Therefore, when the temperature is fixed, the scalar density of
the nucleon increases with the number density of the nucleon, so the
screening masses of $\sigma$ meson and $\omega$ meson increase, while the
effective
mass of the nucleon decreases; When the number density of the nucleon is
fixed, the mean velocity of the nucleon increases with the temperature,
and the scalar density of the nucleon decreases, so the screening masses of
both $\sigma$ meson and $\omega$ meson decrease, while the effective mass
of the nucleon increases.

\section{Photon in the hot nuclear matter}
\label{Photon}

If the electromagnetic interaction between protons in the nuclear matter
is considered, the Lagrangian density on the electromagnetic
interaction can be written as
\begin{equation}
{\cal L}^{\gamma}_{Int}~=~-e\bar\psi \frac{1+\tau_3}{2}\gamma_\mu A^\mu \psi,
\end{equation}
where $\tau_3$ is the Pauli matrix,
\begin{equation}
\tau_3~=~ \left(
\begin{array}{lr}
1 & 0 \\
0 & -1 \\
\end{array}
\right).
\end{equation}
With the similar method in Sec.~\ref{sect:self}, the effective mass of
the real photon in the hot and dense nuclear matter can be
obtained as
\begin{equation}
\label{eq:photon}
m^\ast_\gamma~=~\sqrt{\frac{e^2 \rho_S^p}{2M^{\ast}_N}},
\end{equation}
with $e^2~=~4\pi \alpha$, and $\alpha$ is the fine structure constant. This
result is same as that at zero temperature except that the scalar
density of protons takes the form at finite temperature\cite{Sun.02,Sun.0206}.

The screening mass of the photon in nuclear matter is equal to the
effective mass of the real photon. The effective
mass of the photon corresponding to different densities when
$T=10MeV$ and different temperatures when
$\rho_N=0.148fm^{-3}$ are displayed in Table 1 and 2,
respectively. From Table 1 and 2, it can be concluded that the
effective mass of the photon in nuclear matter increases when the
density of protons increases, while it decreases when the
temperature of nuclear matter increases.

The value of  scalar density of nucleons is similar to that of the
number density of nucleons in the nuclear matter at zero
temperature, $\rho_S^B~\approx~\rho_B$. In the hot nuclear matter,
the distribution of antinucleons should be considered. Supposing the
phase transition to quark gluon plasma is forbidden, in the state
with high temperatures and zero number densities of protons and
neutrons,
\begin{equation}
\rho_p~=~\rho_n~=~0, ~~~and~~~T>>0.
\end{equation}
According to Eq.~(\ref{eq:vmidu}), $n_F(p)=\bar n_F(p)$, and the
scalar density of protons in Eq.~(\ref{eq:biaol}) is not zero, so
the photon gains an effective mass in the state at finite
temperatures and zero number density of nucleons.

\section{Summary}
\label{Summary}

In this paper, according to Wick's theorem, the second order
self-energy corrections of hadrons in the hot and dense nuclear
matter are calculated in the framework of quantum hadrodynamics 1.
Furthermore, the Feynman rules are summarized and the effective
method on quantum hadrodynamics is generalized to the situation at
finite temperatures and densities. As the strong couplings between
nucleons are considered, the self-consistency of this method is
discussed in the relativistic mean-field approximation.

Debye screening masses of the scalar and vector mesons are calculated
with this method in the relativistic mean-field approximation. These
screening meson masses decrease when the temperature becomes higher, while
increase when the density of the nuclear matter increases. In a conclusion,
the results are different
 from those of thermofield dynamics and Brown-Rho conjecture. Moreover, the
effective masses of the photon and the nucleon in the hot and dense nuclear
matter are discussed.

\newpage

\begin{table} [h]
\caption{The screening masses of $\sigma$ and $\omega$ mesons
$m^\ast_\sigma$, $m^\ast_\omega$, the effective mass of the
nucleon $ M^\ast_N$, the energy per nucleon $E/A$ and the
effective mass of the photon $m_\gamma$ in nuclear matter as functions
 of the nucleon number density $\rho_N$ at the temperature $T=10.0MeV$,
 where
$\rho_N$ is in units of the saturation density of nuclear matter
$\rho_0$, and $\rho_0=0.148fm^{-3}$, and the others are in units of
$MeV$.} \vspace*{0.5cm}
\begin{center}
\begin{tabular}{|r|r|r|r|r|r|}\hline
$\rho_N~~(\rho_0)$  &  $m^\ast_\sigma$  &   $m^\ast_\omega$ &
  $ M^\ast_N$  & $E/A$  &  $m_\gamma$ \\ \hline
  0.200  &   546.96  &  798.80  &  848.72  &    7.40  & 2.45\\
  0.400  &   578.10  &  817.62  &  759.73  &    0.45  & 3.65\\
  0.600  &   614.28  &  840.22  &  672.89  &   -5.29  & 4.73\\
  0.800  &   656.26  &  867.36  &  589.54  &   -9.48  & 5.79\\
  1.000  &   704.44  &  899.59  &  511.66  &  -11.61  & 6.87\\
  1.200  &   758.23  &  936.80  &  441.64  &  -11.08  & 7.98\\
  1.400  &   815.74  &  977.87  &  381.57  &   -7.43  & 9.09\\
  1.600  &   874.12  & 1020.75  &  332.30  &   -0.24  &10.16\\
  1.800  &   930.49  & 1063.15  &  293.26  &   10.80  &11.16\\
  2.000  &   983.56  & 1103.87  &  262.46  &   25.18  &12.07\\ \hline
\end{tabular}
\end{center}
\end{table}

\newpage

\begin{table} [h]
\caption{The screening masses of $\sigma$ and $\omega$ mesons
$m^\ast_\sigma$, $m^\ast_\omega$, the effective mass of the
nucleon $ M^\ast_N$, the energy per nucleon $E/A$ and the
effective mass of the photon $m_\gamma$ in nuclear matter as functions
 of the temperature $T$ at the density of nuclear matter
 $\rho_N=\rho_0=0.148fm^{-3}$.
  All are in units of $MeV$.}
\vspace*{0.5cm}
\begin{center}
\begin{tabular}{|r|r|r|r|r|r|}\hline
   T  &  $m^\ast_\sigma$  &   $m^\ast_\omega$ &
  $ M^\ast_N$  & $E/A$ & $m_\gamma$ \\ \hline
   0.00  &  706.53  &  901.01  &  508.10  &  -15.73  &  6.92 \\
  10.00  &  704.44  &  899.59  &  511.66  &  -11.61  &  6.87 \\
  20.00  &  700.08  &  896.62  &  518.06  &   -1.21  &  6.78 \\
  30.00  &  694.85  &  893.09  &  525.88  &   12.49  &  6.67 \\
  40.00  &  689.52  &  889.49  &  534.05  &   28.12  &  6.55 \\
  50.00  &  684.33  &  886.00  &  542.17  &   45.06  &  6.43 \\
  60.00  &  679.40  &  882.70  &  550.08  &   62.98  &  6.32 \\
  70.00  &  674.73  &  879.58  &  557.71  &   81.68  &  6.22 \\
  80.00  &  670.34  &  876.66  &  565.04  &  101.04  &  6.12 \\
  90.00  &  666.22  &  873.93  &  572.06  &  121.00  &  6.02 \\
 100.00  &  662.37  &  871.38  &  578.72  &  141.58  &  5.93 \\
 110.00  &  658.85  &  869.06  &  584.93  &  163.05  &  5.85 \\
 120.00  &  655.79  &  867.05  &  590.40  &  186.16  &  5.78 \\ \hline
\end{tabular}
\end{center}
\end{table}

\newpage

\leftline{\Large {\bf Figure Captions}}
\parindent = 2 true cm
\parskip 1 cm
\begin{description}

\item[Fig. 1] Feynman diagrams on the second order self-energy of the
meson interaction with nucleons in the hot and dense nuclear matter.
The wave lines denote the scalar or vector
meson, 1 and 2 denote particles of the initial
state, 3 and 4 denote particles of the final state. \\

\item[Fig. 2] Feynman diagrams on the second order self-energy of the
nucleon in the hot and dense nuclear matter. The meanings of denotations
are same as those in Fig.~1. \\

\par

\end{description}

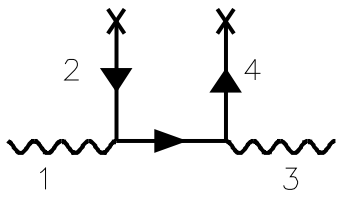
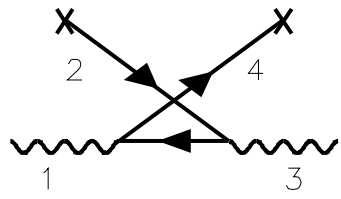
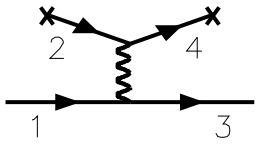
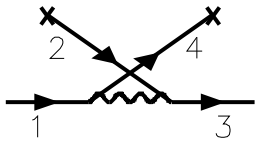
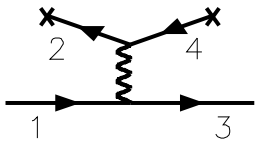
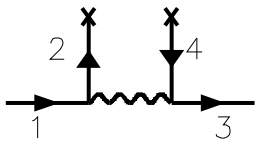
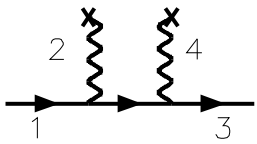
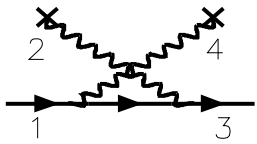
\end{document}